\def \yskip{\penalty-50\vskip3pt plus 3pt minus 2pt}
\def \reference{\par \yskip \noindent \hangindent .4in \hangafter 1}
\def \abc#1#2#3#4 {\reference#1, {\sl#2}, {\bf#3}, #4}
\def \blank {\lower 5pt\hbox to 0.75in{\hrulefill}}

\def \cm{~\rm{cm}}
\def \s{~\rm{s}}
\def \km{~\rm{km}}

\def \K{~\rm{K}}

\def \erg{~\rm{erg}}

\def \yr{~\rm{yr}}
\def \pc{~\rm{pc}}
\def \kpc{~\rm{kpc}}
\def \lesssim{\mathrel{<\kern-1.0em\lower0.9ex\hbox{$\sim$}}}
\def \gtrsim{\mathrel{>\kern-1.0em\lower0.9ex\hbox{$\sim$}}}
\def \BD{BD~+30$^\circ$3639}

\documentstyle[11pt,aasms,tighten,flushrt]{article}
\begin{document}

\title{On the Luminosities and Temperatures of Extended X-ray
Emission from Planetary Nebulae}

\author{Noam Soker $^1$ and Joel H. Kastner$^2$ \\
\small
1. Department of Physics, Oranim, Tivon 36006, 
ISRAEL; soker@physics.technion.ac.il \\
2. Chester F. Carlson Center for Imaging Science, Rochester Institute of 
Technology, 54 Lomb Memorial Dr., Rochester, NY 14623; JHK's
email: jhk@cis.rit.edu }

$$
$$

\centerline {\bf ABSTRACT} 

We examine mechanisms that may explain the luminosities
and relatively low 
temperatures of extended X-ray emission in planetary nebulae.
By building a simple flow structure for the wind from the central
star during the proto, and early, planetary nebulae phase, 
we estimate the temperature of the X-ray emitting gas and its total 
X-ray luminosity.
We conclude that in order to account for the X-ray temperature
and luminosity, both the evolution of the wind from the central star
and the adiabatic cooling of the post-shocked wind's material
must be considered.
The X-ray emitting gas results mainly from shocked wind segments that
were expelled during the early planetary nebulae phase, when the wind
speed was moderate, $\sim 500 \km \s^{-1}$. 
Alternatively, the X-ray emitting gas may result from a
collimated fast wind blown by a companion to the central star. 
Heat conduction and mixing between hot and cool regions
are likely to occur in some cases and may determine the
detailed X-ray morphology of a nebula,
but are not required to explain the basic properties of the X-ray
emitting gas.

\bigskip


\section{INTRODUCTION}

Recent observations of planetary nebulae (PNs) with the
Chandra X-ray Observatory (CXO; \BD, Kastner et al.\ 2000;
NGC 7027, Kastner, Vrtilek, \& Soker 2001a; NGC 6543, Chu et
al.\ 2001) and XMM-Newton (NGC 7009; Guerrero, Gruendl, \&
Chu 2002) have resulted in the discovery of diffuse X-ray
emission in these objects.\footnote{CXO observations have
also led to the discoveries of central point sources with
X-ray emission harder than that of the hot WD in some PNs
(Guerrero et al.\ 2001; Chu et al.\ 2001); the X-rays in
this case are possibly emitted by a magnetically-active main
sequence companion (Guerrero et al.\ 2001; Soker \& Kastner 2002).}
Such extended X-ray emission in
PNs was expected based on the presence of a fast wind driven
by the central star, as the interaction between this fast
wind and slower-moving material, ejected when the
central star was still an asymptotic giant branch (AGB) star, should
lead to energetic shocks.  The same wind-wind interaction
plays a central role in shaping the PN (e.g., Frank 1999,
and references therein). Hence, the processes behind the
extended X-ray emission are tied to the shaping processes of
PNs (Kastner et al.\ 2003), although not all shaping
processes will lead to X-ray emission. 

Subarcsecond imaging by the CXO demonstrates that the
X-ray-emitting regions of some PNs are quite asymmetric.  A
detailed study of the asymmetry of extended X-ray emission
in these PNs was conducted by Kastner et al.\ (2003), where
it was shown that part of the asymmetry of the X-ray
emitting gas results from extinction.  Kastner et al.\
(2003) argue, however, that extinction by itself cannot
account for all of the observed X-ray asymmetries and structural
irregularities, and other processes should be considered.

The CXO and XMM-Newton observations also confirm the results
of previous (ROSAT and ASCA) observations, which indicated that
the temperatures of shock-generated X-ray emission
are a factor $2-10$ lower than predicted by simple energy
conservation arguments.  The X-ray
observations therefore point out the importance of mechanisms
that can moderate the temperature of the shocked gas.  Several such
mechanisms have been proposed previously, most notably,
(1) mixing of nebular and fast wind gas and (2) conduction
fronts in the presence of magnetic fields.  Substantial
mixing of the shocked, fast white dwarf wind with nebular
material would lower the temperature and increase the
density of X-ray emitting material, as observed (Chu \& Ho
1995; Arnaud, Borkowski, \& Harrington 1996, hereafter ABH).  This
mechanism was invoked by Chu et al.\ (2001) as an
explanation for both the low temperature and abundance
anomalies apparent in the spectrum of extended X-ray
emission from NGC 6543. Such mixing might plausibly explain
``clumpy'' X-ray emission morphologies. Alternatively, heat
conduction can result in gas temperatures much lower than
that of the shocked fast wind and, given the presence of
magnetic fields that limit the speed and direction of
conduction fronts, can cause strong departures from spherical symmetry
(Soker 1994; Zhekov \& Perinotto 1996; Zhekov \& Myasnikov 2000).

In the present paper we examine these and other models with
potential to explain the observed temperatures and
luminosities of extended X-ray emission from PNs.
In particular, we consider the effects of adiabatic expansion of
the shocked gas, as well as the potential roles of a wind
driven during post-AGB evolutionary phases and a collimated fast wind (CFW)
blown by a companion to the central star. 
The number of PNs observed thus far justifies a short summary of the
recent observations, which we present
in $\S 2$. In $\S 3$ we discuss potential mechanisms 
underlying the relatively low temperature 
of extended X-ray emission from PNs.
In $\S 4$ we explore, via simple analytical estimates, the emission
from the shocked fast wind material, including its adiabatic 
expansion and evolution with time, assuming that the wind is
generated during the post-AGB phase of the PN progenitor.
We consider these to be the main, but not sole,
processes responsible for the X-ray emitting properties of most PNs.
A summary of our main findings is in $\S 5$. 

\section{PROPERTIES OF PLANETARY NEBULAE IMAGED BY CXO AND XMM}

In Table 1, we summarize the X-ray and other relevant
properties derived for PNs that have been imaged thus far by
CXO and XMM-Newton with the goal of detecting and resolving their
X-ray emission. We include M 1-16, which was observed by
CXO in March 2002 but not detected (Kastner et al., in
preparation). Listed values of X-ray luminosity, taken from
the recent literature, rely on previously published distance
estimates for all sources except \BD, for reasons we now describe.

\subsection{\BD\ and NGC 7027: distances and nebular properties}

Assuming a distance of $d=2.68 \kpc$ to \BD,
Leuenhagen, Hamann, \& Jeffery (1996) find
the central star luminosity and radius to be
$5 \times 10^4 L_\odot$ and
$3.3 R_\odot$, respectively, and the fast wind mass loss rate
to be $\dot M_{f} = 1.3 \times 10^{-5} M_\odot \yr ^{-1}$. 
Given such a high luminosity, the central star most
likely would be the descendant of a massive main sequence
star, and the object's status as a PN would be in doubt.
However, for a distance of $2.68$ kpc, \BD\  would also be $230
\pc$ from the galactic plane, quite high for a massive progenitor.
In addition, the escape velocity from a star with
a mass of $1 M_\odot$ and a radius of $3.3 R_\odot$ is $340 \km \s^{-1}$,
or only about half the observed wind speed; 
thus, it seems that a more compact object is required. Thus,
as noted in Kastner et al.\ (2000),
we conclude the distance to \BD\ is more nearly $\sim1$ kpc.
This estimate is similar to the estimate of 1.5 kpc adopted
by ABH, but is more consistent with estimates summarized in the
Strasbourg-ESO PN Catalogue (Acker et al.\ 1992).
A distance of 1 kpc rather than 2.68 kpc to \BD\ suggests
a much smaller mass loss rate (Table 1),
more compatible with the star's post-AGB, young PN status. The central star
luminosity would be $\sim7000 L_\odot$ at a distance of 1.0 kpc.

The likelihood that \BD\ lies only $\sim1$ kpc from Earth is supported
by its many similarities to NGC 7027, whose distance is
much less uncertain (the 
widely accepted value is $880\pm150$ pc, based on a 
proper motion study of its expanding shell at radio
wavelengths; Masson 1989). Given this distance, and the
(very high) temperature inferred for the central star
($T_\ast \approx 2\times10^5$ K), near-infrared
HST imaging suggests a central star luminosity of $\sim8000
L_\odot$ and radius $0.07 R_\odot$ (Latter et al.\
2000). The very similar luminosities of NGC 7027 and \BD,
assuming a distance of 1.0 kpc for the latter, serve as strong
evidence that these objects had progenitors of similar initial mass
(3--4 $M_\odot$; Bl\"ocker 1995). Many of the differences between
the two nebulae therefore may be due to different viewing angles (Kastner et 
al.\ 2001a) and a few
hundred years of evolution (with NGC 7027 being the
``older'' PN, according to theoretical evolutionary tracks
in Bl\"ocker 1995). 

\subsection{X-ray Emission: Energetics and Temperatures}

From Table 1 it can be seen that the typical total (unabsorbed) X-ray 
luminosities of PNs with detected extended X-ray emission is 
$L_x \simeq 10^{32} \erg \s^{-1}$ (0.3-3.0 keV). 
The kinetic power of the fast wind is
$\dot E_f= \dot M_f v_f^2$ /2, where $\dot M_f$ is the mass loss
rate into the fast wind, and $v_f$ is the fast wind speed.
Thus Table 1 suggests, for a typical X-ray-emitting PN,
\begin{equation}
L_x/\dot E_f \simeq 10^{-2}
\left( \frac { \dot M_f}{3 \times 10^{-8} M_\odot \yr^{-1}} \right)^{-1}
\left( \frac { v_f}{1000 \km \s^{-1}} \right)^{-2}.
\end{equation}
Since the fast wind lasts for at least several hundred years,
the total kinetic energy supplied by the fast wind from the central
star is more than that radiated in the X-ray band during the PN lifetime.

As noted previously, the temperature
of extended, X-ray emitting gas in all PNs detected thus far
by CXO or XMM-Newton ($\sim1-3\times10^6$ K; Table 1) is
substantially lower than 
the expected post-shock temperature, given the speeds of
fast winds from hot, compact PN central stars.
For such (adiabatic shock) conditions, simple arguments 
suggest $T_x \simeq 1.4 \times 10^7 (v_f/1000 \km \s^{-1})^2$,
where $v_f$ is the fast wind speed.
In the following sections we explore the reasons underlying this
apparent discrepancy.

% ============================== TABLE ===============================
%\begin{center}
\begin{table}

Table 1: Properties of Planetary Nebulae Observed by CXO and XMM

\bigskip
\begin{tabular}{|l|c|c|c|c|c|c|}
\hline
 PN & $L_x$  & $T_x$ & $v_f$   & $\dot M_f$ & dynamical age   & morphology \\
&$10^{32}~$erg~s$^{-1}$ &$10^6$~K &km~s$^{-1}$ &$10^{-8} M_\odot \yr^{-1}$ 
    &years & \\
\hline
NGC 7027 & 1.3 & 3 & ? & ? & $\sim 600^{\rm M}$ & EE \\
PN G084.9-03.4&&&&&& \\
\hline
BD +30$^\circ$3639& 1.6 & 3 & $700^{\rm L}$ & $\lesssim 100$ 
          & $\sim 700 $& B pole-on  \\
PN G064.7+05.0&&&&&& or EE?\\
\hline
M1-16 & $<0.02$ & ... & 350$^{\rm CS}$ & ? & $\sim 800^{\rm S}$ & B \\
PN G 226.7+05.6&&&&&& \\
\hline
NGC 6543 & 1.0 & 1.7 & 1,750$^{\rm P}$ & $4^{\rm P}$ &$\sim 1000^{\rm MS}$ 
     & EE\\
PN G096.4+29.9 &&&&&& \\
\hline
NGC 7009 & 0.3 & 1.8& $2800^{\rm C}$ & $0.3 ^{\rm C}$ & $\sim 1700$ & E \\
PN G037.7-34.5&&&&&&   \\
\hline 
NGC 7293 &  $$ & ... & ... & no wind$^{\rm PP}$ & 10000$^{\rm Y}$ & EE  \\
PN G036.1-57.1& & & &     & &\\
\hline
\end{tabular}

\footnotesize
\bigskip

Notes: In the morphology column, `B', `E', and `EE' stand for bipolar,
elliptical, and extreme elliptical, respectively.  
Sources for the X-ray data are:
NGC 7009 (Guerrero et al.\ 2002); BD +30$^\circ$3639 (Kastner et al.\ 2000);
NGC 7027 (Kastner et al.\ 2001a); NGC 7293 (Guerrero et al.\ 2001);
NGC 6543 (Chu et al.\ 2001); M1-16 (Kastner et al.\, in preparation).
Other sources are:
(C) Cerruti-Sola \& Perinotto (1989); (CS) Corradi \& Schwarz (1993);
(L) Leuenhagen et al.\ (1996);
(M) Masson (1989); (MS) Miranda \& Solf (1992);
(P) Perinotto, et al. (1989); (PP) Patriarchi \& Perinotto (1991);
(S) Sahai et al.\ (1994); (Y) Young et al.\ (1999).

\normalsize
\end{table}
%\end{center}
% ======================== END TABLE ===============================

% ===========================================================
\section{CANDIDATE PROCESSES FOR MODERATING X-RAY TEMPERATURE}
% ===========================================================

We now consider six processes that might produce (relatively) low
temperature X-ray-emitting gas.
These processes have been described previously (e.g., ABH),
but we discuss them here in a more coherent way.
We argue that one or more of five of these independent
processes can act together to moderate the X-ray emission
temperature.
In \S 4 we present a detailed study of a particularly promising
model, i.e., the action of a post-AGB wind.
Future observations and calculations will determine the most
important processes in specific systems.

{\bf (1) Cold electrons.} In a collisionless shock the ions are much 
hotter than the electrons (e.g., Laming 2001).  
Coulomb equilibration and plasma instabilities 
will then heat up the electrons. 
For conditions appropriate to young PNs, Coulomb
equilibration by itself will 
heat electrons in a relatively short time. 
Taking a mass loss rate of $10^{-7} M_\odot \yr^{-1}$, a preshock velocity
of $1000 \km \s^{-1}$, and a shock position of $2 \times 10^{16} \cm$
from the central star, we find from equation (3) of Laming (2001) that the 
electrons will be heated up to $\sim 10^7 \K$ in $\sim 100 \yr$. 
Plasma instabilities will make the equilibration time much shorter. 
Therefore, departure from thermal equilibrium is unlikely as a general
explanation for the (relatively) cool X-ray-emitting plasma observed in PNs. 

{\bf (2) Adiabatic expansion.} 
This mechanism may be important for a fast wind which is shocked at early
stages, close to the central star.
Because at early phases of PN evolution
the mass loss rate is higher than at later times,
a fast wind at small radii is shocked to a high density.
This shocked gas expands and cools adiabatically by a large factor 
as the nebula expands, and at late times it
forms a thin, and relatively dense and cool, region in the outer part
of the inner hot bubble (Chevalier \& Imamura 1983; hereafter CI83).
This region emits strongly at lower X-ray temperatures, with
the emission concentrated just within the dense AGB wind.
Therefore, the presence of strong X-ray emission just interior
to the former AGB wind (the dense PN shell, or rim) does not necessarily
imply that there is heat conduction and/or mixing of hot bubble
gas with cold nebular gas in that region. 
Note that relatively dense and cool X-ray emitting gas in the
outer part of the hot bubble forms even if the fast wind has
a speed of $\sim 2000 \km \s^{-1}$ from the beginning
(CI83). In reality, the early slower and denser fast wind makes this region
much more prominent than that seen in figure (3) of CI83, but
for the same reason it cools on a short time scale.
Observationally, the X-ray surface brightness distributions
of NGC 6543  (Chu et al.\ 2001) and \BD\ (Kastner
et al.\ 2003) indicate that
the brightest emission comes from a region close to the dense shell
(rim), with weaker emission from the inner region. The
emission from the inner region of \BD\ is also softer
than the emission from the region adjacent to the optical
rim (Kastner et al.\ 2003), however, suggesting that, at
least for this PN, we must also
consider the evolution of the fast wind (\S 4).

{\bf (3) Mixing of hot and nebular gas.}
This process, which was discussed by Chu \& Ho (1995) for A30 and
ABH for BD $+30^\circ 3639$, is likely to take 
place when there are many small clumps of optically bright 
gas in the nebular interior.
Gas from these clumps, at $\sim 10^4 \K$, is mixed with the fast wind,
thereby lowering its temperature. 
ABH argue that the mixing ratio required to form the 
cooler X-ray emitting gas should result in elevated oxygen
abundance. In contrast, analyses of X-ray spectra of \BD\
(ABH; Kastner et al.\ 2000) and NGC 6543 (Chu et al.\ 2001)
suggest that the X-ray-emitting gas is, if anything,
somewhat depleted in oxygen. Thus, while mixing may occur to
some extent in certain nebulae, it is unlikely to explain in
general the relatively low X-ray emission temperatures of PNs.

{\bf (4) Heat conduction.} 
 It has been noticed previously that heat conduction 
can form PN X-ray emission regions
having temperatures much lower than that of the shocked fast wind
(Soker 1994; Zhekov \& Perinotto 1996), and can cause departures from
spherical symmetry (Soker 1994; Zhekov \& Myasnikov 2000). 
The heat-conduction front goes through three stages:
evaporative, quasi-static, and condensation. 
 During the first two stages, the intermediate temperature region
is inside the initially cold gas, while in the third phase it
is in the initially hot gas (Borkowski, Balbus, \& Fristrom 1990).
 This may explain certain aspects of the observations that
are not easily explained by, e.g., mixing of nebular and fast wind
material.
 For example, in the frame of the heat conduction front
model, the depleted abundance of oxygen indicated by the
X-ray spectra of \BD\ suggests that the X-ray-emitting
conduction front in this PN is in one of the first two phases;
i.e., it is located in the initially cold (nebular) gas.
 The interface during the early stages
(the evaporation and quasi-static stages) is at a temperature
of $T_I \gtrsim 0.4 T_h$, where $T_h$ is the temperature of the
hot gas (Borkowski et al.\ 1990). 
 Therefore, in the heat conduction model, the emitting gas
at $T_x \sim 3 \times 10^6 \K$ is the gas ejected during the
AGB phase, consistent with the result that its oxygen
abundance is much lower than the present abundance of
the central Wolf-Rayet star.
 In NGC 6543, on the other hand, the abundances are those of
the central fast wind (Chu et al.\ 2001), despite the rather
low temperature of the X-ray emitting gas (Chu et al.)
and the very high speed of the wind (1750 km s$^{-1}$;
Perinotto et al.\ 1989). 
 In the present context, this is explained as a heat conduction 
front in the condensation  phase; that is, the intermediate temperature
region lies within the initially hot (fast wind) gas.
 A similar situation appears to characterize NGC 7009 (Guerrero
et al.\ 2002).
 Hence, invoking the heat conduction model, we would
conclude that the heat conduction fronts in NGC 6543 and NGC
7009 are more evolved than those in \BD. 

{\bf (5) Post-AGB wind.}
To produce shocks with characteristic temperatures $T_x \approx 2-3
\times 10^6 \K$, the wind speed should be $\sim 400-500 \km \s^{-1}$,
where the exact value depends on the adiabatic expansion of
the post-shock material.
Such a ``moderate speed'' wind likely arises in the central
star's late post-AGB, and early PN, phases, as the star
traverses the H-R diagram (e.g., Kastner et al.\ 2001b). 
ABH have proposed that a moderate-speed wind, generated at the
post-AGB phase, is the source of the X-ray emitting gas in \BD.  
Whereas the shocked moderate-speed gas cools very quickly, the
high speed gas cools slowly and emits at a low rate per unit volume.
Therefore --- as we demonstrate below ---
even if a faster ($\sim 1000 \km \s^{-1}$) central star wind
is present, the dominant 
X-ray emission would still come from gas originating in a
more moderate-speed, post-AGB wind driven at
$\sim 500 \km \s^{-1}$.
 
{\bf (6) Collimated fast wind from a companion.}
A main sequence companion outside the progenitor AGB star may accrete
from the AGB wind, forming an accretion disk and driving
jets or a CFW at speeds of $\sim 200-600 \km \s^{-1}$.
  Such a mechanism, described in detail by Soker \& Rapapport (2000), 
would be similar to that thought responsible for 
X-ray emission apparently associated with protostellar
outflows (e.g., Pravdo et al.\ 2001; Favata et al.\ 2002) and
for the collimated X-ray-emitting jets in the symbiotic binary
system R Aqr (Kellogg, Pedelty \&  Lyon 2001).
Interestingly, the temperatures of the X-ray-emitting plasma in
these systems are similar to those in X-ray-emitting young PNs,
i.e., $\sim10^6$ K.
The observations of R Aqr are perhaps most relevant, as 
symbiotic binaries are frequently associated with PNs displaying
pronounced bipolar structure.
If such a companion-driven, collimated wind
is responsible for the X-ray emission in young
PNs, then the X-ray emission should be confined to the
polar regions, as is apparently the case for NGC 7027, NGC 6543, and,
possibly, \BD.
In this scenario, furthermore, the X-ray emission properties
of most elliptical PNs, which have no collimated winds, should
be very different from those of bipolar PNs (although some
elliptical PNs may have been shaped by main sequence
companions which generated a CFW; Soker 2001).  

% ===========================================================
\section{POST-AGB WIND AS THE MAIN SOURCE OF THE X-RAY EMISSION}
% ===========================================================

We now consider in more detail shocks due to a moderate-speed,
post-AGB wind, and the adiabatic expansion of the shocked gas.
The shocked gas may originate in a CFW from a companion
to the central star as well. 
 We build a simple flow structure to derive the typical
temperature and X-ray luminosity of observed PNs (Table 1).
We assume the following:
\begin{enumerate}
\item The boundary between the shocked fast wind and the AGB wind,
i.e., the contact discontinuity, expands at a constant speed,
$v_{\rm sh}$.
Its location is at $r_{\rm sh} = v_{\rm sh} t$, where $t$ is the
age of the fast wind.
This is the location of the inner boundary of the
dense AGB shell (i.e., the nebular rim) that is compressed
by the shocked fast wind. 
From observations $v_{\rm sh} \simeq 20 \km \s^{-1}$.
\item The fast wind is shocked (at the reverse shock) at a radius of
$r_f = \eta r_{\rm sh}$, where $\eta \simeq 0.3$ is a constant.
Assumptions (1) and (2) are based on the self-similar solution
of CI83, which assumes constant wind properties.
 This is not the case here, since in post-AGB stars the fast
wind evolves; specifically, its velocity increases while the
stellar mass loss rate decreases.
{{{ In addition, ionization, which is not incorporated in
the self-similar solution of CI83, may play a significant
role and, in particular, can significantly change the value of
$v_{\rm sh}$ (e.g., Marten \& Sch\"onberner 1991; Mellema 1994;
Marigo et al.\ 2001; Steffen \& Sch\"onberner 2002).
Under some conditions the boundary can move much slower than the
undisturbed AGB wind (e.g., Steffen \& Sch\"onberner 2002)
or may even move inward, such that $v_{\rm sh}<0$
(Figs.\ 5-7 of Mellema 1994; Fig.\ 9 of Marigo et al.\ 2001).
However, over a long time interval the contact discontinuity expands
(e.g., Fig.\ 11 of Marigo et al.\ 2001).
Furthermore, the general shapes of the PNs found to have
extended X-ray emission are not spherical.
Hence, even if the contact discontinuity in the narrower
equatorial region moves inward, we still expect the
discontinuity to move outward in the polar directions;
therefore the hot bubble volume increases, as assumed here. }}}
\item The pressure inside the hot bubble is constant with location,
but not with time, and is given by $P=0.85 \rho_f v_f^2$ (CI83).
Here $v_f$ is the velocity of the fast wind, 
$\rho_f=\dot M_f/(4 \pi r_f^2 v_f)$ is the density of the fast
wind just before hitting the reverse shock, and $\dot M_f$ is the
mass loss rate characterizing the fast wind.
\item The cooling function in the temperature range
$2 \times 10^5 \K \lesssim T \lesssim 2 \times 10^7 \K$ is
$\Lambda=10^{-22} (T/10^6K)^{-1/2} \erg \cm^3 \s^{-1}$, which is
an approximation to figure 5 of Gaetz, Edgar, \& Chevalier (1988).
\end{enumerate}
The cooling rate per unit volume per unit time is
$\Lambda n_e n_p$, where $n_e$ and $n_p$ are the
electron and proton density, respectively.
As the gas cools its cooling time decreases quickly.
Taking this into account, the cooling time is 
\begin{equation}
\tau_{\rm cool} \simeq
\frac {n k T} {\Lambda n_e n_p} = \frac {P}{\Lambda n_e n_p},
\end{equation}
where $n$ is the total number density.
Substituting for the pressure the value from assumption (3) above,
gives 
\begin {equation}
\tau_{\rm cool} \simeq
200  % 188.3
\left( \frac {T} {10^6 \K}  \right)^{5/2}
\left( \frac {v_{\rm sh}} {20 \km \s^{-1}}  \right)^2
\left( \frac {t} {500 \yr}  \right)^2
\left( \frac {\dot M_f v_f} {3 \times 10^{-5} M_\odot \yr^{-1} \km \s^{-1}}
   \right)^{-1}
\left( \frac {\eta} {0.3}  \right)^2 \yr.
\end{equation}
 Parcels of gas for which the cooling time is shorter than the
age of the hot bubble effectively are removed from the hot bubble;
{{{ that is, these parcels of gas cool fast, and no longer contribute 
to the pressure and energy content of the hot bubble. }}}
 This condition, $\tau_{\rm cool} \lesssim t$, implies
that the minimum temperature of the gas in the hot bubble is
\begin{equation}
T_{\rm min} \simeq 1.5 \times 10^6  % 1.47x10^6
\left( \frac {v_{\rm sh}} {20 \km \s^{-1}}  \right)^{-0.8}
\left( \frac {t} {500 \yr}  \right)^{-0.4}
\left( \frac {\dot M_f v_f} {3 \times 10^{-5} M_\odot \yr^{-1} \km \s^{-1}}
   \right)^{0.4}
\left( \frac {\eta} {0.3}  \right)^{-0.8}.
\end{equation}           

Using the above result for cooling time, we can estimate the luminosity
of a shocked segment of wind.
 We look at a time period $\Delta t$ during which the fast wind
has a speed $v_f$ and therefore is shocked to a temperature of 
$kT=3 \mu m_p v_f^2/16$, where $\mu m_p$ is the mean mass
per particle in the (ionized) shocked wind.
The kinetic energy of the fast wind is
$\Delta E_k = 0.5 \dot M_f v_f^2 \Delta t = 
\dot M_f v_f (4 k T /3 \mu m_p )^{1/2} \Delta t $.
The approximate X-ray luminosity of the shocked gas during this period is 
\begin {eqnarray}
L_x \simeq \frac {\Delta E_k}{\tau_{\rm cool}}
\simeq 10^{33}  %1.34
\left( \frac {T} {10^6 \K}  \right)^{-2}
\left( \frac {v_{\rm sh}} {20 \km \s^{-1}} \right)^{-2}
\left( \frac {t} {500 \yr}  \right)^{-2}
\left( \frac {\Delta t} {100 \yr}  \right) \\ \nonumber
\left( \frac {\dot M_f v_f} {3 \times 10^{-5} M_\odot \yr^{-1} \km \s^{-1}}
   \right)^{2}
\left( \frac {\eta} {0.3}  \right)^{-2} \erg \s^{-1}.
\end{eqnarray}

The following things should be noted in regards to the last equation.
\begin{enumerate}
\item There is a dependency between some variables.
If, for example,  $\dot M_f v_f$ is larger, i.e., a stronger fast wind,
then the acceleration of the dense remnant of the AGB wind is more
efficient, and $v_{\rm sh}$ is faster.
\item The variation of $\eta$ is more complex (CI83).
With time, as the fast wind's mass loss rate decreases and
its velocity increases, the reverse shock moves inward and $\eta$ decreases;
hence, at early stages, $\eta$ can be larger than $0.3$.
{{{ We note that, in these equations, $\eta$ actually appears
as $\eta v_{\rm sh}$.
 Including the effects of ionization --- principally, 
the resulting reduced (or even negative) velocity of the
contact discontinuity (e.g., Mellema 1994;
Marigo et al. 2001) --- will further complicate the evolution.
However, as stated earlier, for our present analytical study
the complex shapes of the X-ray-emitting PNs and the other
uncertainties, e.g., in the evolution of the fast wind,
don't warrant inclusion of ionization effects. }}}
\item Increasing $\dot M_f v_f$ will increase $L_x$. Hence, a
PN with a strong core wind --- e.g., a PN with a [WC] central star ---
should have stronger X-ray emission than a PN with a
weaker central star wind, at a given evolutionary
time. However, $L_x$ does not depend as sensitively on
$(\dot M_f v_f)$ as is suggested by Eq.\ 5, since
increasing $\dot M_f v_f$ also leads to
more efficient cooling (Eq.\ 3) and, hence, more shocked fast
wind gas will be removed from the hot bubble.
\item If the fast wind were constant at $v_f \gtrsim 1000 \km
\s^{-1}$, as is the case during a non-negligible portion of the
evolutionary lifetimes of some PNs,
the temperature of shocked wind gas should be $T\gtrsim 10^7 \K$.
Thus, if some other mechanism --- such as adiabatic
expansion, heat conduction, or mixing --- is not taken into account, then 
the X-ray luminosity would be $L_x \ll 10^{32} \erg \s^{-1}$,
much less than the observed values.
\end{enumerate}
For present purposes we consider only the effect of
adiabatic expansion of the shocked fast 
wind gas on $T_x$ and $L_x$; the reader is referred to
references in points 3 and 4 in section 3 for detailed discussions
of the mixing and heat conduction processes, respectively.

After a fast wind segment is shocked at time $t_{\rm shock}$
to a temperature $T_{\rm shock}$, it
expands slowly with the hot bubble and the nebula, hence it
cools adiabatically (CI83).
For a constant fast wind, the density in the hot bubble
decreases as $t^{-2}$, and the assumption of adiabatic cooling 
implies $T=T_{\rm shock} (t/t_{\rm shock})^{-4/3}$.
The thermal energy of the gas (per unit mass) decreases as $T$ decreases,
while the cooling time goes as $T^{5/2}$ (Eq.\ 3).
Using these dependencies in Eq.\ 5, we can derive a
very crude expression for the X-ray luminosity of a fast wind
segment that was blown during a time $\Delta t$ and was shocked at time
$t_{\rm shock}$, i.e., 
\begin {eqnarray}
L_x \simeq 10^{32}
\left( \frac {T_{\rm shock}} {3 \times 10^6 \K}  \right)^{-2}
\left( \frac {v_{\rm sh}} {20 \km \s^{-1}} \right)^{-2}
\left( \frac {t_{\rm shock}} {500 \yr}  \right)^{-2}
\left( \frac {\Delta t} {100 \yr}  \right) \\ \nonumber
\left( \frac {\dot M_f v_f} {3 \times 10^{-5} M_\odot \yr^{-1} \km \s^{-1}}
   \right)^{2}
\left( \frac {\eta} {0.3}  \right)^{-2} \erg \s^{-1}.
\end{eqnarray}
Although the time $t$ does not appear in the last equation,
there is an implicit time dependance.
As the nebula gets older, earlier wind segments cool and droop
from the X-ray emitting material (eqs. 3 and 4).
Therefore, as the time $t$ increases, the dominate X-ray emitting gas
comes from later wind segments, i.e., higher values of $t_{\rm shock}$,
and $L_x$ decreases, as indeed demonstrated by equation (5).
Even when adiabatic cooling is taken into account,
a fast wind with $v_f > 1000 \km \s^{-1}$ and 
$T_{\rm shock} \gtrsim 1.5 \times 10^7 \K (t/1000 \yr)^2$
fails to account for the observed X-ray luminosities of
X-ray-emitting PNs, as the shocked bubble gas is still too hot.
We must consider the wind evolution as well.

The evolution of the fast wind during the post-AGB and early PN 
phases is poorly known and likely depends
sensitively on parameters that are exceedingly difficult to
ascertain, such as the mass and multiplicity of the central star.
Nevertheless we take, as a crude approximation for these
early evolutionary phases, 
$v_f \simeq 1000 (t/1000 \yr )$ (Sch\"onberner \& Steffen 2000), which gives 
$T_{\rm shock} \simeq 1.4 \times 10^7 \K (t/1000 \yr)^2$. 
Substituting this relation in equation (6), gives  
\begin {eqnarray}
L_x \sim 10^{32}
\left( \frac {v_{\rm sh}} {20 \km \s^{-1}} \right)^{-2}
\left( \frac {t_{\rm shock}} {500 \yr}  \right)^{-6}
\left( \frac {\Delta t} {100 \yr}  \right) \\ \nonumber
\left( \frac {\dot M_f v_f} {3 \times 10^{-5} M_\odot \yr^{-1} \km \s^{-1}}
   \right)^{2}
\left( \frac {\eta} {0.3}  \right)^{-2} \erg \s^{-1} .
\end{eqnarray}

\section{CONCLUSIONS AND SUMMARY}

Equations (4)-(7), though crude, and the other results and discussions
in $\S 3$ and 4, lead to several fundamental conclusions,
which we now describe. We also consider the
interplay between the various processes that can moderate
the temperature of diffuse X-ray emission from PNs, as it is
likely that five of six of the processes described in \S 3
(i.e., all processes aside from
non-equilibrium electron temperatures) are important in
governing the onset and evolution of X-ray emission in PNs.
\begin{enumerate}
\item To satisfactorily 
account for the observed X-ray luminosities and temperatures of PNs,
the evolution of the post-AGB wind must be considered, together
with its adiabatic cooling. When both effects are considered,
the X-ray luminosity and temperature can be explained in
terms of the action of a late post-AGB (or early PN phase) 
wind --- with a speed lower ($\sim 500 \km \s^{-1}$) than
those of evolved PNs ($\sim 1000 \km \s^{-1}$) ---
which had time to cool adiabatically.
This critical wind phase spans a time period of $\sim 100-200 \yr$. 
Wind segments from earlier, rapid-mass-loss, post-AGB phases
are slower and undergo more adiabatic 
cooling and, hence, are effectively removed from the hot
bubble, according to equation (4). 
\item Jets, or a collimated fast wind (CFW), driven by a main sequence 
companion at the termination of the AGB, should have much
the same properties as the post-AGB wind from the primary.  
Hence, in a bipolar PN that was shaped by a CFW, a substantial
or even dominant fraction of the X-ray emitting gas may come from
a CFW (or jets) blown by a companion. Such may be the case
for \BD\ and NGC 7027. Although neither
nebula is known to harbor a binary central star,
both nebulae display fast, collimated outflows 
(Bachiller et al.\ 2000; Cox et al.\ 2002) that suggest
the presence of companions. On the other hand, the
nondetection of M1-16 (Table 1) ---
which {\it is} thought to possess a CFW blown by a companion (Schwarz 1992;
Sahai et al.\ 1994) --- demonstrates that a CFW doesn't
necessarily produce diffuse X-ray emission. 
We speculate that the lack of X-ray emitting gas in M 1-16 can be
ascribed to strong binary interactions, which have yielded a
PN core with low mass and luminosity (Schwarz 1992;
Sahai et al.\ 1994) and, therefore, a weak wind.
\item Shocked fast wind episodes that occur later will become hotter
than the shocked post-AGB wind or CFW and
should display far lower total luminosity, by about
an order of magnitude. Hence, we still expect some (very faint) 
diffuse X-ray emission from shocked fast wind gas at $T>3\times 10^6 K$. 
\item Heat conduction from the hot bubble to the dense nebular shell,
i.e., the remnant of the AGB wind, was proposed in the past as a mechanism 
to reduce the temperature of the X-ray emitting gas and to shape the 
PN (Soker 1994; Zhekov \& Perinotto 1996; Zhekov \& Myasnikov 2000).
A weak magnetic field can inhibit heat conduction, if the field lines
do not run from the hot bubble to the AGB wind remnant. 
Here we raise the possibility that heat conduction occurs between 
the hot bubble and the segment of the shocked fast wind that has already
cooled radiatively to $\sim 10^4 \K$. 
The effects will be similar to those resulting from heat conduction 
from shocked gas to the slow wind, but it is more likely that there will be
magnetic field lines connecting the already-cool shocked fast wind to
the shocked fast wind material that remains at high temperature. 
The same argument holds for mixing of cool and hot gas, which was suggested
to occur between the AGB wind remnant and the hot bubble
(Chu \& Ho 1995; ABH; Chu et al.\ 2001); here
we raise the possibility that such mixing more likely occurs between
already-cool and still-hot shocked fast wind gas.
\item The X-ray luminosity steeply decreases with time, as can
be seen from equations (5) and (7); 
note that in equation (7) as time $t$ proceeds, the X-ray will come
mostly from shocked gas at later $t_{\rm shock}$.
The decrease becomes more dramatic after the factor $\dot
M_f v_f$ starts to decrease, and the fast wind ceases. 
 Therefore, we expect any diffuse X-ray emission from
old PNs (like NGC 7293) to be weak, and perhaps undetectable.
\item The presently X-ray-emitting gas was ejected several hundred years
ago. It may have had a different composition than the present
composition of the wind.
This may explain the apparent abundance contrast between the
X-ray-emitting gas and the present central star wind (ABH).
As mention previously, heat conduction may also
account for that observation.
\end{enumerate}

The next steps, beyond the initial treatment presented here,
should be taken by gasdynamical simulations, first in
spherical geometry to study the post-AGB wind, then in 2
dimensions to study the role of a CFW.  These simulations
should consider the time evolution of the shocked post-AGB
wind and its X-ray properties, so as highlight potential
distinctions between the X-ray properties of gas originating
in a post-AGB wind versus that originating in a CFW blown by
a possible companion. Even given such simulations, however,
it may not be possible to distinguish between these
alternatives, since all of the PNs observed thus far by CXO
and XMM are bipolar or extreme elliptical PNs which may have
been shaped by a CFW. Observations of elliptical and
spherical PNs, which are not expected to have been shaped by
a CFW, will be required to break this degeneracy.

\acknowledgements{

J.H.K. acknowledges support for this research
provided by NASA/CXO grant GO0--1067X to RIT. 
N.S. acknowledges support from the US-Israel
Binational Science Foundation. }

\end{document}